\documentclass[10pt,conference]{IEEEtran}
\usepackage[printonlyused]{acronym}

\usepackage{caption}
\usepackage{}
\usepackage{subcaption}
\usepackage{bm}
\usepackage{enumitem}
\usepackage{array, makecell}
\usepackage{graphicx}
\def\BibTeX{{\rm B\kern-.05em{\sc i\kern-.025em b}\kern-.08em
    T\kern-.1667em\lower.7ex\hbox{E}\kern-.125emX}}
{}

\makeatletter
\renewcommand{\fnum@figure}{Figure \thefigure}
\makeatother

\begin{document}

\title{Experimental Assessment of Neural 3D Reconstruction for Small UAV-based Applications}
\vspace{-7mm}
\author{
\IEEEauthorblockN{Genís Castillo Gómez-Raya\IEEEauthorrefmark{1}, Álmos Veres-Vitályos\IEEEauthorrefmark{1}, Filip Lemic\IEEEauthorrefmark{1}\IEEEauthorrefmark{2}\textsuperscript{\textsection}, Pablo Royo\IEEEauthorrefmark{3}, \\Mario Montagud\IEEEauthorrefmark{1}, Sergi Fernández\IEEEauthorrefmark{1}, Sergi Abadal\IEEEauthorrefmark{3}, Xavier Costa-P\'erez\IEEEauthorrefmark{1}\IEEEauthorrefmark{5}}
\vspace{1mm}
\IEEEauthorblockA{\IEEEauthorrefmark{1}i2CAT Foundation, Spain}
\IEEEauthorblockA{\IEEEauthorrefmark{2}Faculty of Electrical Engineering and Computing, University of Zagreb, Croatia}
\IEEEauthorblockA{\IEEEauthorrefmark{3}Universitat Politècnica de Catalunya, Spain}
\IEEEauthorblockA{\IEEEauthorrefmark{5}NEC Laboratories Europe GmbH, Germany and ICREA, Spain\\
Email: filip.lemic@i2cat.net}
\vspace{-6mm}
}

% \author{
% \IEEEauthorblockN{Genís Castillo Gómez-Raya, Álmos Veres-Vitályos, Filip Lemic, Pablo Royo Chic, Mario Montagud, \\
% Sergi Fernández, Sergi Abadal, Xavier Costa-Pérez}
% \vspace{-7mm}
% }

\maketitle

\begingroup\renewcommand\thefootnote{\textsection}
\footnotetext{Corresponding Author.}
\endgroup

\begin{abstract}
The increasing miniaturization of \acp{UAV} has expanded their deployment potential to indoor and hard-to-reach areas. 
However, this trend introduces distinct challenges, particularly in terms of flight dynamics and power consumption, which limit the \acp{UAV}' autonomy and mission capabilities. 
This paper presents a novel approach to overcoming these limitations by integrating \ac{GenAI} with small \ac{UAV} systems for fine-grained \ac{3D} digital reconstruction of small static objects. 
Specifically, we design, implement, and evaluate an \ac{GenAI}-based pipeline that leverages advanced models, i.e., Instant-ngp, Nerfacto, and Splatfacto, to improve the quality of 3D reconstructions using images of the object captured by a fleet of small UAVs.
We assess the performance of the considered models using various imagery and pointcloud metrics, comparing them against the baseline \ac{SfM} algorithm. 
The experimental results demonstrate that the \ac{GenAI}-enhanced pipeline significantly improves reconstruction quality, making it feasible for small \acp{UAV} to support high-precision 3D mapping and anomaly detection in constrained environments.
In more general terms, our results highlight the potential of \ac{GenAI} in advancing the capabilities of miniaturized \ac{UAV} systems.
\end{abstract}

%!TEX root = ieee_pimrc_main.tex

\acrodef{KDE}{Kernel Density Estimation}
\acrodef{UAV}{Unmanned Aerial Vehicle}
\acrodef{SfM}{Structure from Motion}
\acrodef{3D}{3-Dimensional}
\acrodef{2D}{2-Dimensional}
\acrodef{LPD}{Loco Positioning Deck}
\acrodef{LPS}{Loco Positioning System}
\acrodef{UWB}{Ultra Wide-Band}
\acrodef{TDoA}{Time Difference of Arrival}
\acrodef{VR}{Virtual Reality}
\acrodef{RGB}{Red, Green and Blue}
\acrodef{CLI}{Command Line Interface}
\acrodef{NeRF}{Neural Radiance Fields}
\acrodef{TWR}{Two-Way Ranging}
\acrodef{TDoA}{Time Difference of Arrival}
\acrodef{ULP}{Ultra Low Power}
\acrodef{IoT}{Internet of Things}
\acrodef{IO}{Input-Output}
\acrodef{PCB}{Printed Circuit Board}
\acrodef{GS}{Gaussian Splatting}
\acrodef{CAGR}{Compound Annual Growth Rate}
\acrodef{MLP}{Multilayer Perceptron}
\acrodef{AI}[AI]{Artificial Intelligence}
\acrodef{GenAI}[N3DR]{Neural 3D Reconstruction}
\acrodef{PSNR}[PSNR]{Peak Signal-to-Noise Ratio}
\acrodef{SSIM}[SSIM]{Structural Similarity Index}
\acrodef{LPIPS}[LPIPS]{Learned Perceptual Image Patch Similarity}
\acrodef{HD}{Hausdorff Distance}
\acrodef{WD}{Wasserstein Distance}
\acrodef{MVS}{Multi-View Stereo}
\acrodef{RANSAC}{Random Sample Consensus}
\acrodef{SotA}{State-of-the-Art}
\acrodef{ADC}{Adaptive Density Control}
\acrodef{DTR}{Differentiable Tile Rasterizer}
\acrodef{PSR}{Poisson Surface Reconstruction}
\acrodef{xR}{eXtended Reality}
\acrodef{RT}{Real Time}
\acrodef{RTK}{Real-Time Kinematic}
\acrodef{GPS}{Global Positioning System}
\acrodef{MSE}{Mean Squared Error}
\acrodef{NN}{Neural Network}
%!TEX root = ieee_pimrc_main.tex

%%%%%%%%%%%%%%%%%%%%%%%%%%%%%%%%%%%%%%%%%%%%%%%%%%%%%%%%%%%%%%%%%%%%%%%%%%%%%%%%%%%%%%%%%
%%%%%%%%%%%%%%%%%%%%%%%%%%%%%%%%%%%%%%%%%%%%%%%%%%%%%%%%%%%%%%%%%%%%%%%%%%%%%%%%%%%%%%%%%
%% Introduction
%%%%%%%%%%%%%%%%%%%%%%%%%%%%%%%%%%%%%%%%%%%%%%%%%%%%%%%%%%%%%%%%%%%%%%%%%%%%%%%%%%%%%%%%%
%%%%%%%%%%%%%%%%%%%%%%%%%%%%%%%%%%%%%%%%%%%%%%%%%%%%%%%%%%%%%%%%%%%%%%%%%%%%%%%%%%%%%%%%%
% \vspace{-1mm}
\section{Introduction}

The use of \acfp{UAV} has rapidly increased and is expected to continue growing~\cite{nex2022uav}. 
As \acp{UAV} become smaller, their potential applications can be expanded in the direction of operating in more diverse and confined environments such as indoors and hard-to-reach areas. 
However, this miniaturization also brings distinct challenges, particularly regarding flight dynamics and power consumption~\cite{floreano2015science}. 
Smaller \acp{UAV} encounter issues such as reduced power density in electromagnetic motors, transmission inefficiency due to increased friction in bearings and gears, and greater viscous losses due to decreased Reynolds numbers. 
Furthermore, flight modes like hovering become more energy-intensive and difficult to control as \ac{UAV} sizes decrease. 
These scaling issues fundamentally limit the autonomy of \acp{UAV}, thereby restricting their ability to carry out complex missions~\cite{floreano2015science}.

Beyond \ac{UAV} miniaturization challenges, \acf{3D} reconstruction and anomaly detection are key enablers for applications like infrastructure inspection and precision manufacturing. 
Traditional methods struggle with small UAVs and fine-scale objects due to sensor limitations such as low-resolution cameras, limited field of view, and motion blur, resulting in incomplete reconstructions with occlusions, shadows, perspective and scale errors, etc. 
Systems integrating \ac{GenAI} with miniaturized UAVs are currently lacking, yet they are envisioned to offer a solution to these challenges by enhancing accuracy, mitigating artifacts, and enabling high-fidelity reconstructions.
Aligned with this vision, we introduce an \ac{GenAI}-based pipeline for the 3D digital reconstruction of small static objects using images captured by a single or fleet of small \acp{UAV}. Advances in technology have enabled even small \acp{UAV} to capture images of static objects, generating datasets that can be processed by \ac{GenAI} models to create 3D object representations.
We employ a nano-\ac{UAV} (i.e., Crazyflie~2.1) system for \ac{3D} reconstruction of a small object, leveraging \ac{GenAI} to enhance the quality of reconstructions. Three state-of-the-art models, i.e., Instant-ngp~\cite{muller2022instant}, Nerfacto~\cite{nerfstudio}, and Splatfacto~\cite{kerbl3Dgaussians}, are evaluated using datasets of images from both single and multi-\ac{UAV} setups.
A comprehensive set of imagery and pointcloud metrics is used to evaluate the resemblance between the reconstructions and a 3D-printed object. 
The results highlight the performance of the \ac{GenAI} models and demonstrates the feasibility of using small \acp{UAV} in combination with \ac{GenAI} for 3D anomaly detection.

\subsection{Main Contributions}

\paragraph{Redefining evaluation methodologies for small-scale 3D reconstruction} As 3D reconstructions shift toward smaller objects and scenes, traditional evaluation methodologies designed for macroscopic-scale reconstructions are starting to lose relevance. Our results reveal that metrics such as \ac{PSNR}, \ac{LPIPS}, and \ac{WD} lack the resolution to capture subtle variations in small-scale objects. This study emphasizes the need for new evaluation frameworks tailored to fine-grained geometric discrepancies. While Splatfacto provides the highest reconstruction accuracy, Nerfacto exhibits more stable performance across different objects, making it preferable for anomaly detection. These findings highlight the necessity of revising evaluation methodologies for N3DR applications in small-scale robotics.

\paragraph{Advancing 3D reconstruction and anomaly detection toward micro-scale} As recent advances in robotics are enabling device and system miniaturization, there is an increasing need for advanced computational techniques that operate effectively at small scales. This paper demonstrates that \ac{GenAI} pipelines leveraging state-of-the-art models can enhance 3D digital reconstruction and anomaly detection using small UAVs. 
By system downsizing, this study contributes toward the long-term vision of wireless robotic materials, which are distributed computational structures capable of self-monitoring and adaptation~\cite{correll2017wireless,lemic2021survey}. 
The evaluation of single- and dual-UAV setups, along with location-aware reconstruction, demonstrates that even resource-constrained UAVs can contribute to fine-grained anomaly detection, paving the way for future applications in ultra-miniaturized robotics.

%!TEX root = ieee_pimrc_main.tex

%%%%%%%%%%%%%%%%%%%%%%%%%%%%%%%%%%%%%%%%%%%%%%%%%%%%%%%%%%%%%%%%%%%%%%%%%%%%%%%%%%%%%%%%%
%%%%%%%%%%%%%%%%%%%%%%%%%%%%%%%%%%%%%%%%%%%%%%%%%%%%%%%%%%%%%%%%%%%%%%%%%%%%%%%%%%%%%%%%%
%% Related Works
%%%%%%%%%%%%%%%%%%%%%%%%%%%%%%%%%%%%%%%%%%%%%%%%%%%%%%%%%%%%%%%%%%%%%%%%%%%%%%%%%%%%%%%%%
%%%%%%%%%%%%%%%%%%%%%%%%%%%%%%%%%%%%%%%%%%%%%%%%%%%%%%%%%%%%%%%%%%%%%%%%%%%%%%%%%%%%%%%%%
\vspace{-1mm}
\section{Related Works}

Photogrammetry is an established technique for extracting shape information from physical objects and scenes~\cite{schenk2005introduction}. 
It is widely used in applications such as topographic mapping, archaeology, architecture, infrastructure maintenance, and gaming~\cite{kovanivc2023review}. 
Photogrammetry involves capturing multiple images from different viewpoints and using them to calculate the \ac{3D} coordinates of an object’s surface. 
This method is particularly effective when combined with \acf{SfM}, which does not require prior knowledge of camera positions, making it more flexible than traditional stereo imaging techniques that mimic human binocular vision~\cite{kamencay2012improved}. 

\begin{figure*}
\centering
\begin{minipage}{.65\textwidth}
  \centering
  \includegraphics[width=0.98\linewidth]{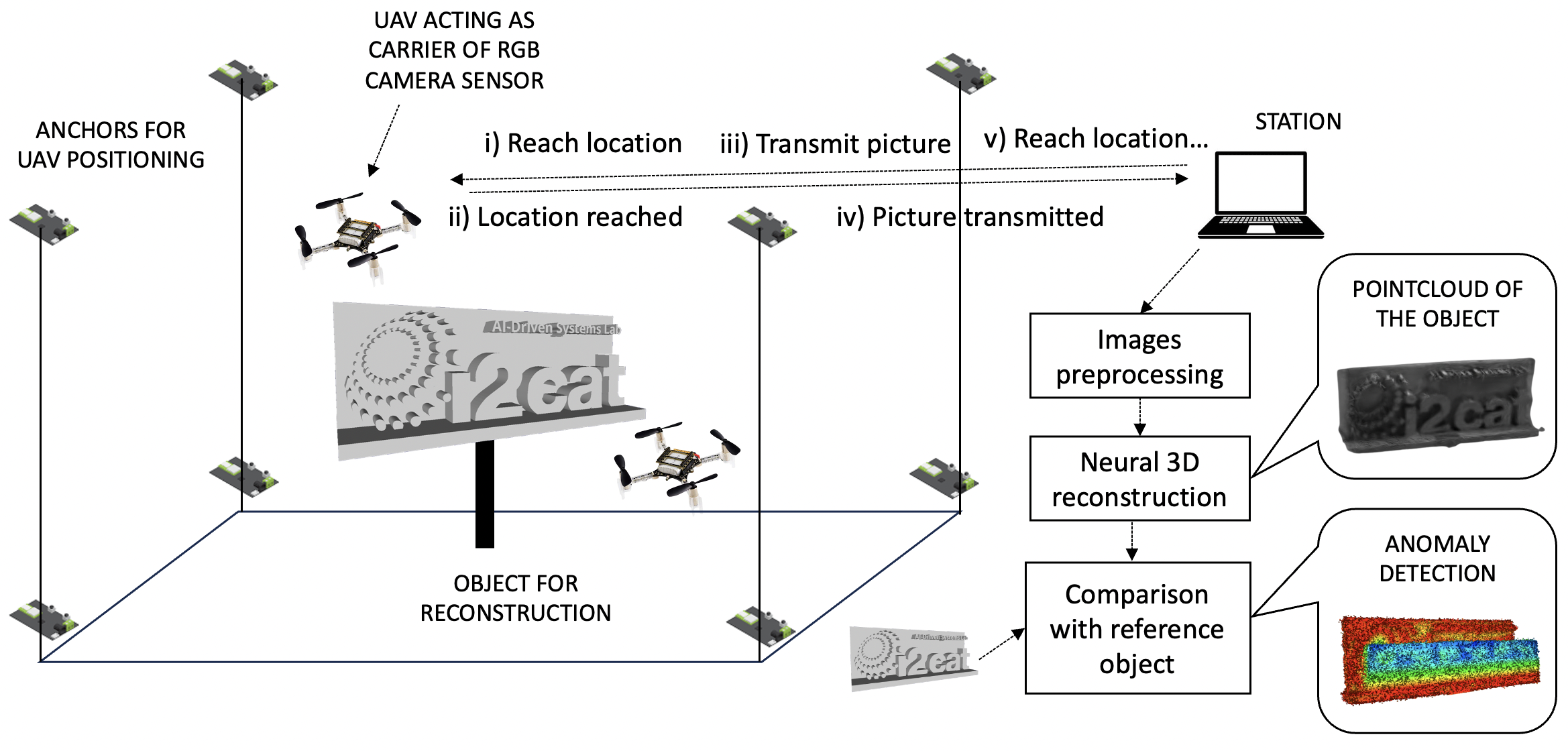}
  \vspace{-4mm}
  \captionof{figure}{Hardware and software setup}
  \label{fig:hardware_setup}
  \vspace{-1mm}
\end{minipage}
\hfil
\begin{minipage}{.18\textwidth}
  \centering
  \includegraphics[width=\linewidth]{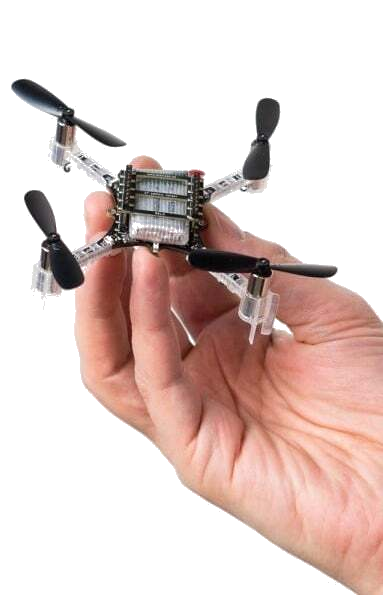}
  \captionof{figure}{Crazyflie~2.1}
  \label{fig:drone}
\end{minipage}%
\vspace{-3mm}
\end{figure*}

\ac{SfM} has been widely adopted due to its robustness and ability to handle large datasets~\cite{kamencay2012improved}. 
It reconstructs \ac{3D} structures by identifying matching features across overlapping images, estimating camera positions, and triangulating points to create a sparse \ac{3D} pointcloud. 
This approach is advantageous as it does not necessitate precise camera positioning or prior knowledge of camera intrinsic parameters, which can instead be estimated directly from the images. 
However, despite its flexibility, \ac{SfM}-based methods struggle with reconstructing fine-grained objects, particularly when working with low-resolution sensors or when image sequences contain occlusions, shadows, or perspective distortions.

\ac{GenAI}-based techniques have emerged as a promising alternative, leveraging deep learning to improve reconstruction accuracy, mitigate artifacts, and enhance fine-scale detail extraction. 
\ac{GenAI} synthesizes novel views of an object or scene by learning an implicit volumetric representation from a set of images~\cite{gao2022nerf,mildenhall2021nerf}. 
Although \ac{GenAI} provides high-fidelity reconstructions, its computational complexity presents a challenge for near-\ac{RT} applications, particularly in resource-constrained platforms such as miniaturized \acp{UAV}. 
Several state-of-the-art \ac{GenAI} models have been developed, each optimizing different aspects of the reconstruction.  

Instant-ngp introduces a multi-resolution hash encoding that accelerates training while maintaining reconstruction accuracy, making it particularly suitable for near-\ac{RT} applications~\cite{muller2022instant}.  
This encoding optimizes feature vectors in each iteration for improved memory efficiency, with performance influenced by feature vector size, hash table dimensions, and resolution.

Nerfacto, an advanced \ac{NeRF} implementation from Nerfstudio~\cite{mildenhall2021nerf,nerfstudio}, refines camera poses via backpropagated loss gradients, addressing errors stemming from inaccurate camera positioning and slight motion artifacts. 
A piecewise sampler distributes samples efficiently, while a proposal sampler focuses on key scene areas, using an \ac{MLP}-based density function with multi-resolution hash encoding. 
The density field determines object presence, while the \ac{RGB} field specifies color. After training, these fields enable \ac{NeRF} visualization and scene rendering, producing a pointcloud from ray samples and generating a mesh through Poisson surface reconstruction.

Splatfacto~\cite{kerbl3Dgaussians}, based on \ac{GS}, employs CUDA-accelerated differentiable rasterization to optimize scene representations, significantly reducing the computational cost associated with NeRF-style reconstructions.  
It initializes \ac{3D} Gaussians based on sparse \ac{SfM} points, projects them into \ac{2D} space using camera parameters, and refines density using adaptive control.  
The dynamic texture refinement process aligns projections with input images, generating gradients that iteratively refine Gaussian shape and color.  
The final optimized Gaussians enable scene rendering using a Gaussian-based method.

By comparing these techniques, this study evaluates the feasibility of \ac{GenAI}-driven reconstruction for small \acp{UAV}, considering key factors such as accuracy, computational efficiency, and robustness to imaging artifacts.  
Although traditional \ac{SfM}-based methods remain relevant in some scenarios, \ac{GenAI} approaches offer superior performance, particularly in cases where high-fidelity reconstructions are required despite low sensor resolution and constrained computational resources.  
The integration of these models with lightweight \ac{UAV}-based imaging systems has the potential to significantly enhance autonomous \ac{3D} mapping, anomaly detection, and inspection capabilities in constrained environments.

%!TEX root = ieee_pimrc_main.tex

%%%%%%%%%%%%%%%%%%%%%%%%%%%%%%%%%%%%%%%%%%%%%%%%%%%%%%%%%%%%%%%%%%%%%%%%%%%%%%%%%%%%%%%%%
%%%%%%%%%%%%%%%%%%%%%%%%%%%%%%%%%%%%%%%%%%%%%%%%%%%%%%%%%%%%%%%%%%%%%%%%%%%%%%%%%%%%%%%%%
%% Energy Model 
%%%%%%%%%%%%%%%%%%%%%%%%%%%%%%%%%%%%%%%%%%%%%%%%%%%%%%%%%%%%%%%%%%%%%%%%%%%%%%%%%%%%%%%%%
%%%%%%%%%%%%%%%%%%%%%%%%%%%%%%%%%%%%%%%%%%%%%%%%%%%%%%%%%%%%%%%%%%%%%%%%%%%%%%%%%%%%%%%%%
% \vspace{-2mm}
\section{Evaluation Methodology}
\label{sec:methodology_results}

\subsection{Evaluation Setup}
We use a small-\ac{UAV}-based experimentation infrastructure designed for autonomous 3D reconstruction of small static objects, as shown in Figure~\ref{fig:hardware_setup}. 
The infrastructure employs lightweight Crazyflie 2.1 UAVs (cf., Figure~\ref{fig:drone}), each weighing less than 100~grams, integrated into an open-source pipeline for near-\ac{RT} 3D reconstruction.  

The system leverages the \ac{LPS} for UAV localization. The LPS consists of \ac{UWB} anchors distributed across the environment and a tag attached to the UAV. After the anchors are positioned and calibrated, the system localizes the UAV with the average accuracy of less than 10~cm~\cite{mendes2022small}, achieved through \ac{TDoA}-based multilateration.

The UAVs are equipped with cameras that capture black-and-white images from various angles, which are wirelessly transmitted to a ground station for 3D model generation. The near-RT reconstruction capability allows dynamic adjustments to the UAVs’ flight paths based on the \ac{RT} quality of the 3D model, optimizing both coverage and accuracy. 
Both single- and dual-UAV configurations are considered, operating in two modes: \textit{baseline 3D reconstruction}, where only the images are used for reconstruction, and \textit{location-aware 3D reconstruction}, where positional data from the \ac{LPS} is combined with the images to improve accuracy. 

\subsection{Evaluation Methodology}
\label{sec:methodology}

The results of each experiment are the render, mesh, and pointcloud visualizations of a reconstructed object.
The pointclouds are used in evaluation due to their wide utilization and availability of programmatic frameworks for the metrics' calculation. 
The reference object is indicated in Figure~\ref{fig:hardware_setup}, which was 3D-printed for the evaluation. 
The resulting 3D-printed object features the size of 54.7$\times$20.3$\times$20.9~cm\textsuperscript{3}, with the letters and engravings having a depth of 4~cm.

The alignment between a reconstructed pointcloud and a reference was performed as follows.
Pointcloud orientation and scaling consistency were achieved by computing scaling factors from bounding box ratios and employing fast global registration and principal component analysis for orientation correction.  Alignment was refined through global registration techniques based on feature matching and the iterative closest point algorithm. Post-alignment, virtual cameras were generated uniformly around the object and directed towards its center, to render images from both the reference and reconstructed pointclouds. Performance metrics were subsequently calculated directly from the pointclouds or from comparisons with the rendered images.
The utilized metrics are:
\begin{itemize}[leftmargin=*]
\item\textbf{\ac{PSNR}} is the ratio between the maximum possible signal power and the noise affecting its quality, used to compare the \ac{MSE} between images and renders.
\item\textbf{\ac{SSIM}} indicates similarity between images and renders, with $1$ being perfect similarity. This metric compares luminance, contrast, and structure differences across the image samples.
\item\textbf{\ac{LPIPS}} is the perceptual similarity between the images and the 3D render.
\item\textbf{\ac{HD}} captures the overall deviation of a reconstruction as the maximum distance between the points in the reference and reconstructed pointclouds.
\item\textbf{\ac{WD}} calculates the minimum “work” required to transform a reconstructed pointcloud into the reference, measuring the cost of transporting points between the two.
\item\textbf{Reconstruction latency} refers to the total time taken to calculate a full reconstruction from the input images for the \ac{SfM} approach. The latency for the \ac{GenAI} approaches will be captured as the time required for generating the final pointcloud representation, which combines the model training and pointcloud rendering.
\end{itemize}

To enable anomaly detection, we establish a quantitative comparison between the reconstructed object and a known reference.  
The anomaly detection process consists of computing the baseline metric on interest for a reconstructed object that does not feature an anomaly, comparing it against a reference object without anomalies.  
This is followed by computing the metric for a reconstructed object that contains an anomaly, using the same reference object without anomalies.  
Finally, we calculate the difference in the metric for a reconstructed object with and without the anomaly, which quantifies the structural difference introduced by the anomaly.  
This approach allows for both quantitative and spatial anomaly detection. The magnitude of the difference provides an indication of how pronounced the anomaly is, while spatial anomaly localization can be achieved by analyzing the differences between the reconstructed and reference pointclouds. 

\subsection{Image Preprocessing}

\begin{figure}[!t]
\centering
\begin{subfigure}{0.2368\textwidth}
    \centering
    \includegraphics[width=\linewidth]{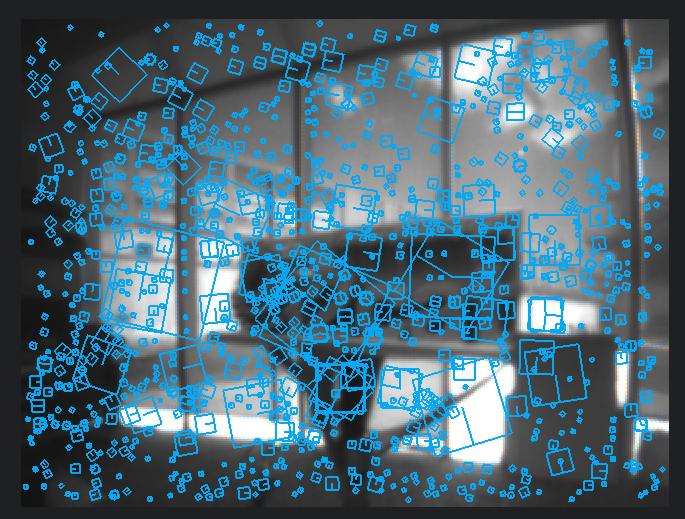}
    \caption{1206 features on original}
\end{subfigure}
\begin{subfigure}{0.235\textwidth}
    \centering
    \includegraphics[width=\linewidth]{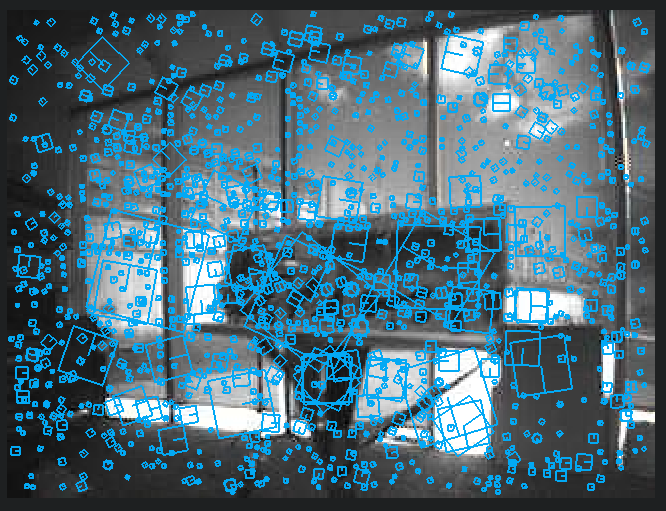}
    \caption{1513 features on processed}
\end{subfigure}
\caption{Improvements due to image preprocessing}
\label{fig:image_preprocessing}
\vspace{-4mm}
\end{figure}

The images are processed through sharpening and brightness adjustments, which were empirically optimized to improve the reconstruction quality, as shown in Figure~\ref{fig:image_preprocessing}. 
As illustrated, the edited images capture more distinct features around the object compared to the originals, leading to more robust feature detection. 
This is evident in the matches between consecutive frames that are more consistent in the edited images, as shown by the clustering and detection of feature points. The grouping of images, based on the UAV's yaw and timestamp, ensures coherent surface and structure reconstruction, further enhancing the fidelity of the 3D model.

\subsection{Hyperparameter Tuning}

\begin{figure}[!t]
\centering
    \includegraphics[width=0.85\linewidth]{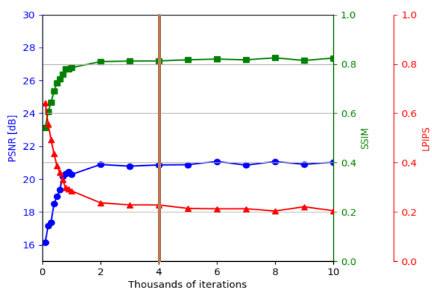}
    \vspace{-1mm}
\caption{Hyperparameter optimization for Nerfacto}
\label{fig:hyperparameter_tuning}
\vspace{-4mm} 
\end{figure}

In the considered \ac{GenAI} models, the hyperparameters optimized include the allowed maximum iterations for both the feature matching algorithm and the localizer estimator during the \ac{SfM} process.
These parameters were uniformly set across all tests to ensure consistent comparisons.
During feature matching, the \ac{RANSAC} algorithm~\cite{derpanis2010overview} is used to filter out incorrect feature matches that do not conform to a consistent transformation. 
\ac{RANSAC} works by randomly sampling subsets of feature matches and fitting a model to them, followed by evaluating how many other matches are consistent with the model. 
By adjusting the number of allowed iterations, the computational time can be reduced, but this may result in noisier reconstructions and potentially lower quality.
The number of iterations for the localizer estimator and the feature matching algorithm was set empirically. 
An example is depicted in Figure~\ref{fig:hyperparameter_tuning} for Nerfacto, with the brown vertical line indicating the selected number of iterations. Our aim was to strike a balance between reconstruction accuracy and execution time.

There is a discrepancy between the metrics' values presented in the hyperparameter optimization in Figure~\ref{fig:hyperparameter_tuning} and the final results. 
During hyperparameter tuning, the PSNR, SSIM, and LPIPS metrics are calculated based on the rendered images of the scene, directly comparing the quality of the rendered output to a reference. This image-based evaluation typically shows higher metrics due to the nature of 2D evaluation.
In contrast, the final results are calculated from the pointclouds of the scene rather than the rendered images, and demonstrate lower values (cf., Tables~\ref{tab:results_reconstruction} and~\ref{tab:results_anomaly_detection}). This is due to pointcloud-based evaluation being more stringent, focusing on the accuracy of the 3D geometry and structure, which is more sensitive to small deviations. 
This decision was made with the understanding that the hyperparameter tuning process focused on providing a near-RT reconstruction capability, prioritizing efficiency. While extended training would yield slightly better results, the iteration limits were tuned to deliver acceptable reconstruction quality within a constrained execution time. 

\begin{table*}[!t]
  \caption{Performance of different approaches for 3D reconstruction of an object without an anomaly}
  % \vspace{-1mm}
  \label{tab:results_reconstruction}
    \centering
    \small
    \begin{tabular}{l c c c c c c c c}
    \hline
    \textbf{Approach} & \textbf{PSNR~[dB]} &  \textbf{SSIM} &  \textbf{LPIPS} & \textbf{HD} &  \textbf{WD} &  \textbf{Latency~[s]}  & \makecell{\textbf{\# images} \\\textbf{taken}} & \makecell{\textbf{\# images} \\\textbf{used}} \\ \hline
    \multicolumn{9}{c}{\textbf{Single-UAV system - Baseline}} \\ 
    \ac{SfM}        & 7.405& 0.972$\pm$0.006 & 0.050$\pm$0.007&0.077 &0.0285 & 271& 233& 232 \\ 
    Instant-ngp &7.750&0.956$\pm$0.008&0.052$\pm$0.009& 0.039& 0.0152& 757 &233 &232 \\ 
    Nerfacto        & 7.565&0.960$\pm$0.007&0.042$\pm$0.007& 0.027& 0.0164 & 444&233 &232  \\  
    Splatfacto       &7.730&0.971$\pm$0.007&0.049$\pm$0.012& 0.033&0.0171 &  302&233 &232 \\ \hline
    \multicolumn{9}{c}{\textbf{Single-UAV system - Location-aware solution}} \\ 
    Instant-ngp &7.989& 0.957$\pm$0.008&0.052$\pm$0.009& 0.039&0.0188 &876 &233 &232 \\ 
    Nerfacto        & 7.710&0.963$\pm$0.006&0.040$\pm$0.008& 0.025&0.0165
    &493 & 233&232 \\ 
    Splatfacto     &7.852&0.969$\pm$0.007&0.044$\pm$0.011& 0.033&0.0140 
    & 314&233 &232 \\ \hline
    \multicolumn{9}{c}{\textbf{Dual-UAV system - Location-aware solution}} \\ 
    Instant-ngp     &7.750&0.957$\pm$0.009&0.050$\pm$0.009& 0.033&0.0156&
    783 &466 &465 \\ 
    Nerfacto       &7.642&0.963$\pm$0.006&0.039$\pm$0.007& 0.025&0.0164&475 &466 &465 \\
    Splatfacto    & 7.666&0.967$\pm$0.008&0.049$\pm$0.009& 0.035&0.0152&302 &466 &465 \\ 
    \hline
    \end{tabular}
\vspace{-1mm}
\end{table*}

\begin{table*}[!t]
  \caption{3D reconstruction performance for an object with anomaly and anomaly detection using HD metric}
  % \vspace{-1mm}
  \label{tab:results_anomaly_detection}
    \centering
    \small
    \begin{tabular}{l c c c c c c c c c}
    \hline
        \textbf{Approach} & \textbf{PSNR~[dB]} &  \textbf{SSIM} & \textbf{LPIPS}&\textbf{HD}&\textbf{WD}&\textbf{Latency~[s]}&\bm{$HD_B$} & \bm{$HD_A$} & \bm{$\Delta HD$} \\ \hline
        \multicolumn{9}{c}{\textbf{Baseline}} \\ 
        \ac{SfM}        &7.195 &0.971$\pm$0.006 &0.060$\pm$0.010 &0.072 &0.0469 & 306& 0.077& 0.092& 0.015\\ 
        Instant-ngp     &7.212&0.952$\pm$0.008&0.058$\pm$0.009& 0.034& 0.0164 &754 &0.039&0.073& 0.034 \\ 
        Nerfacto        &7.319&0.960$\pm$0.007&0.044$\pm$0.007& 0.024&0.0141&465 &0.027&0.062& 0.035 \\  
        Splatfacto     &7.745&0.970$\pm$0.008&0.048$\pm$0.010& 0.039 &0.0177 &302 &0.033&0.070& 0.037\\
         \hline
        \multicolumn{9}{c}{\textbf{Location-aware solution}} \\ 
        Instant-ngp      &7.254&0.955$\pm$0.009&0.054$\pm$0.009&0.034&0.0148& 837& 0.039&0.076&0.037 \\
        Nerfacto        &7.341&0.962$\pm$0.006&0.042$\pm$0.007&0.025 &0.0152&470 &0.025&0.076&0.051 \\  
        Splatfacto     &7.285&0.971$\pm$0.006&0.045$\pm$0.009&0.034&0.0161& 310& 0.033&0.074& 0.041 \\  

        \hline
    \end{tabular}
    \vspace{-2mm}
\end{table*}

\section{Evaluation Results}
\label{sec:evalution_results}

In Table~\ref{tab:results_reconstruction}, we overview the \ac{3D} reconstruction performance in case the object of interest does not feature an anomaly. 
The performance is assessed for the three considered \ac{GenAI} models and the \ac{SfM} baseline along the heterogeneous set of metrics overviewed in Section~\ref{sec:methodology}.
When comparing the single- and dual-UAV setups, it is visible that the inclusion of the second \ac{UAV} does not benefit the reconstruction performance across all metrics and approaches. 
This suggests that a sufficient number of images was collected in a single-UAV setup for reaching the performance limits across approaches, with the number of images as indicated in Table~\ref{tab:results_reconstruction}. 
In other words, the reported performance is constrained by the inherent limitations of \ac{GenAI} for the given set of low quality images. 
For this reason, in the remainder we focus on the single-UAV setup.

The objects generated by \ac{GenAI} are collections of several small pieces which might appear hollow from certain angles. This phenomenon is observable on the results, because the AI-based solutions did not improve or yielded only marginal improvements in terms of \ac{PSNR}, \ac{SSIM}, and \ac{LPIPS} compared to the SfM baseline. This is because the imagery metrics are derived using images rendered from the object with randomly chosen cameras. The pointcloud metrics are based on the inspection of the objects, so they are immune to this effect. 

One can observe that not all metrics fluctuate significantly across the considered approaches. 
This is primarily the case for \ac{PSNR}, \ac{LPIPS}, and \ac{WD} that remain at constant values of respectively $7.7\pm0.2$, $0.05\pm0.01$, and $0.016\pm0.002$ across approaches. 
We argue that the lack of sensitivity is caused by the fact that the object for reconstruction is rather small. 
The majority of the current literature targets the reconstruction of comparably larger objects, which intuitively feature larger reconstitution errors and, therefore, larger difference in the reconstruction when compared to the references. 
Due to the fact that \ac{PSNR}, \ac{LPIPS}, and \ac{WD} lack sensitivity to reflect the differences between reconstruction quality of the assessed approaches, we consider the other metrics as more relevant and focus on them in the remainder of the paper.

\begin{figure*}[!t]
\centering
\begin{subfigure}{0.238\textwidth}
    \centering
    \includegraphics[width=\linewidth]{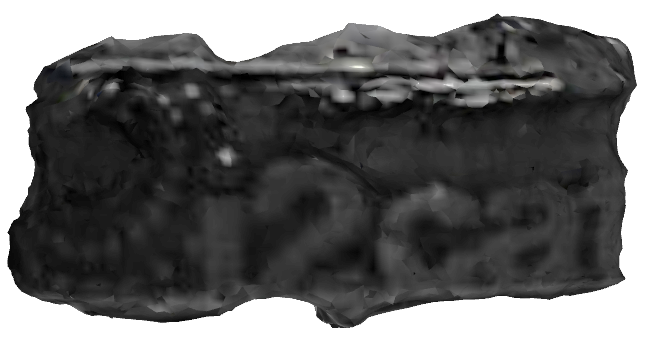}
    \caption{SfM}
\end{subfigure}
\hfill
\begin{subfigure}{0.238\textwidth}
    \centering
    \includegraphics[width=\linewidth]{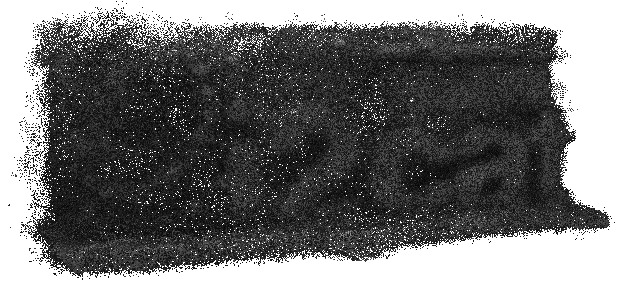}
    \caption{Location-aware instant-ngp}
\end{subfigure}
\hfill
\begin{subfigure}{0.238\textwidth}
    \centering
    \includegraphics[width=\linewidth]{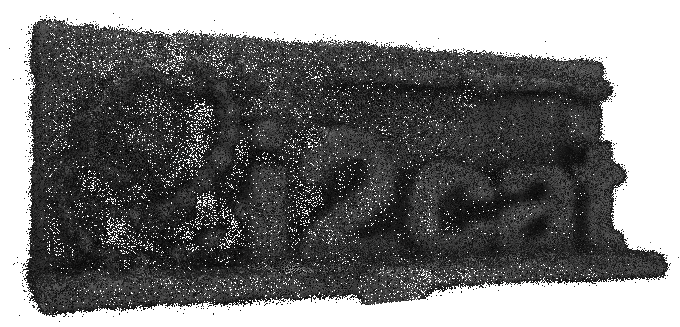}
    \caption{Location-aware Nerfacto}
\end{subfigure}
\hfill
\begin{subfigure}{0.238\textwidth}
    \centering
    \includegraphics[width=\linewidth]{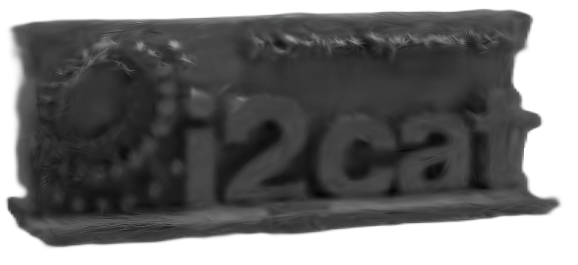}
    \caption{Location-aware Splatfacto}
\end{subfigure}
% \vspace{-1mm}
\caption{3D reconstruction capabilities for different approaches for single-UAV system}
\label{fig:reconstruction_1uav}
\vspace{-2mm}
\end{figure*}

\begin{figure*}[!t]
\centering
\begin{subfigure}{0.238\textwidth}
    \centering
    \includegraphics[width=\linewidth]{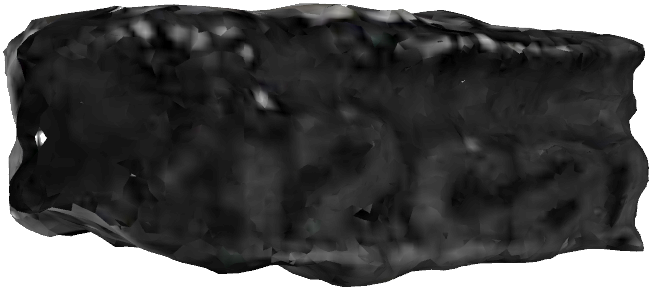}
    \caption{SfM}
\end{subfigure}
\hfill
\begin{subfigure}{0.238\textwidth}
    \centering
    \includegraphics[width=\linewidth]{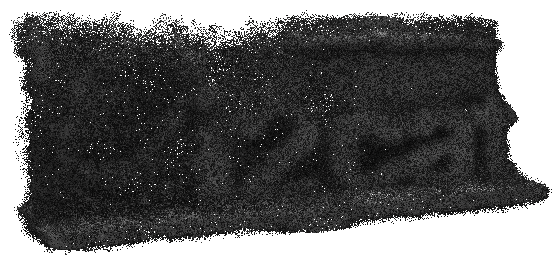}
    \caption{Location-aware instant-ngp}
\end{subfigure}
\hfill
\begin{subfigure}{0.238\textwidth}
    \centering
    \includegraphics[width=\linewidth]{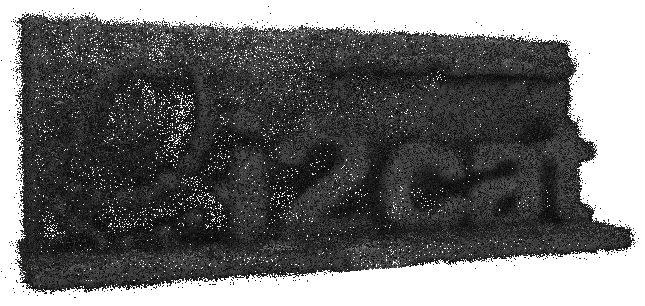}
    \caption{Location-aware Nerfacto}
\end{subfigure}
\hfill
\begin{subfigure}{0.238\textwidth}
    \centering
    \includegraphics[width=\linewidth]{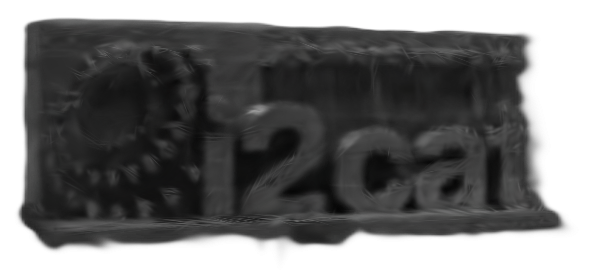}
    \caption{Location-aware Splatfacto}
\end{subfigure}
% \vspace{-1mm}
\caption{3D reconstruction capabilities for different approaches for dual-UAV system}
\label{fig:reconstruction_2uav}
\vspace{-3mm} % Adjust space after the caption if needed
\end{figure*}

The utilization of location coordinates stemming from the \ac{LPS} in addition to the ones obtained through \ac{SfM} (i.e., location-aware solution) does not significantly benefit the reconstruction accuracy. 
The reason for that can be found in the fact that the locations provisioned by the \ac{LPS} coordinate system feature certain errors, with its average value in the range of 10~cm, as reported in~\cite{mendes2022small}. 
Nonetheless, the location-aware approach enhances the reliability of pointcloud generation by offering an additional source for provisioning of camera locations. 
In other words, using \ac{SfM} to obtain coordinates is only feasible if the aligned images contain enough features for accurate camera localization, whereas the location-aware approach allows the \ac{UAV} location coordinates to be used as an alternative source.
This can be observed in Figures~\ref{fig:reconstruction_1uav} and~\ref{fig:reconstruction_2uav} by comparing the baselines and location-aware methods, where it is observable that the utilization of location-awareness increases the density of the resulting pointclouds compared to the corresponding baselines.
Intuitively, the advantages of context-aware approaches are anticipated to be more significant for larger target objects in reconstruction, where the image density relative to the object's size is lower.

\begin{figure*}[!t]
\centering
\begin{subfigure}{0.238\textwidth}
    \centering
    \includegraphics[width=\linewidth]{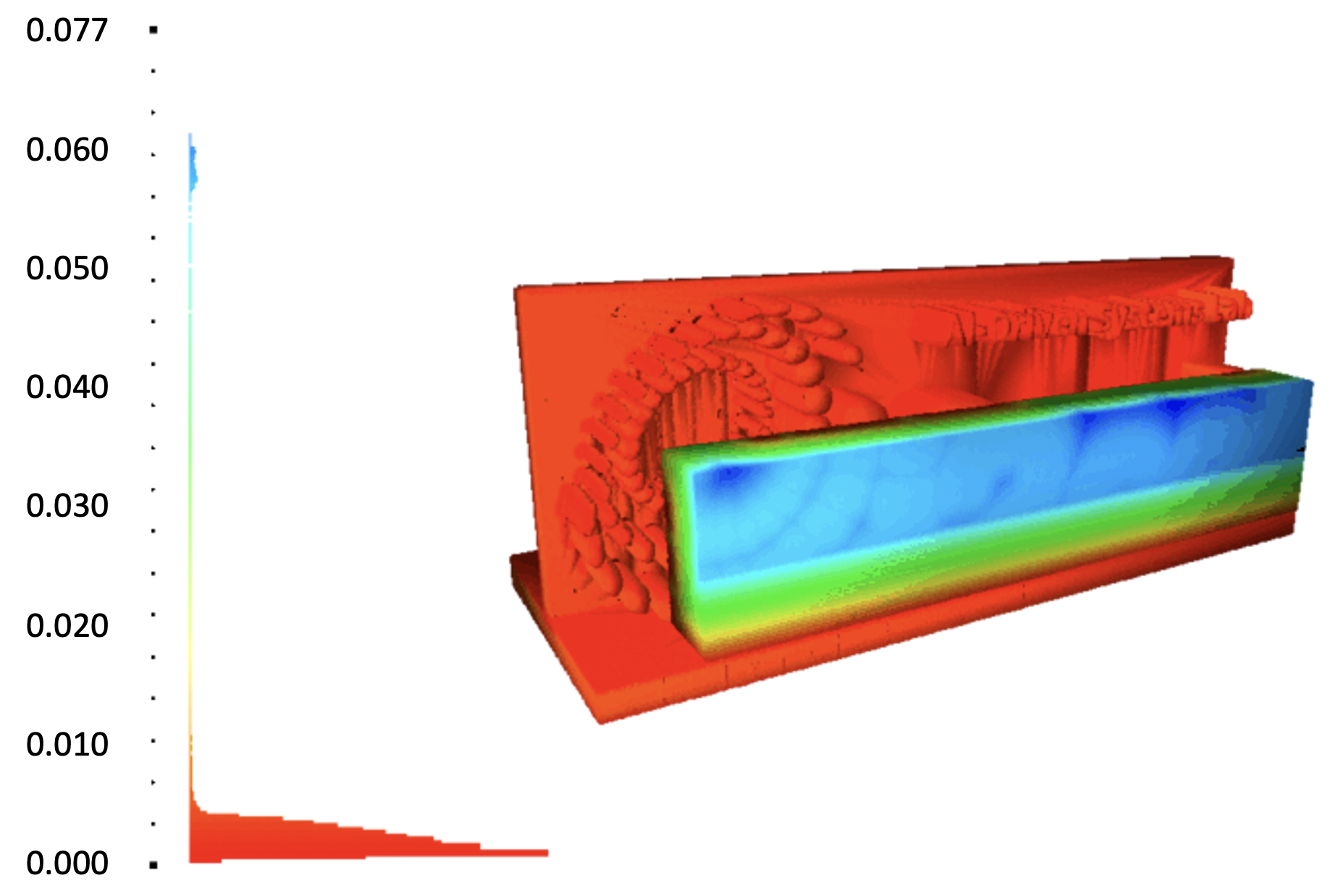}
    \caption{Oracle}
\end{subfigure}
\hfill
\begin{subfigure}{0.238\textwidth}
    \centering
    \includegraphics[width=\linewidth]{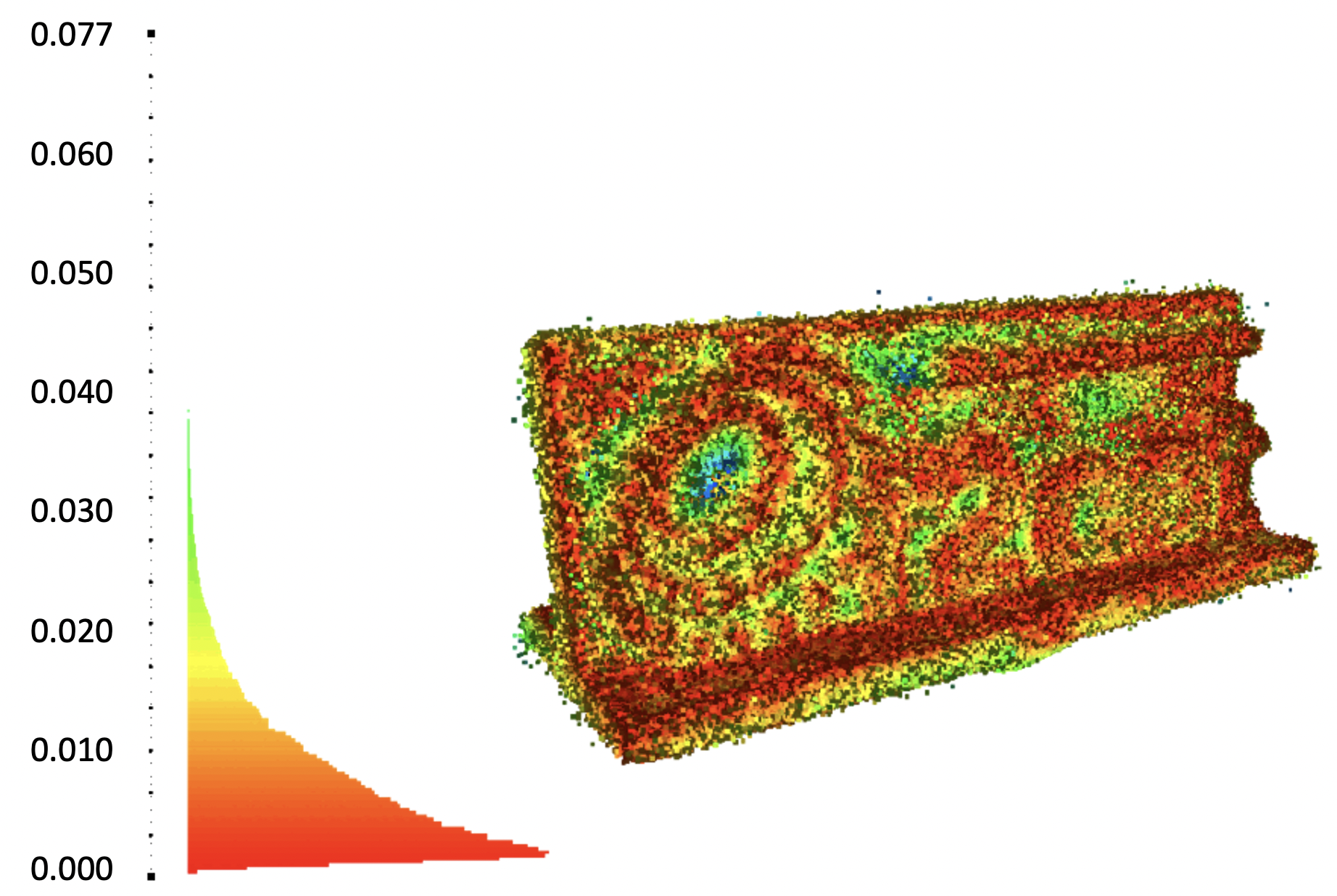}
    \caption{No anomaly detected}
\end{subfigure}
\hfill
\begin{subfigure}{0.238\textwidth}
    \centering
    \includegraphics[width=\linewidth]{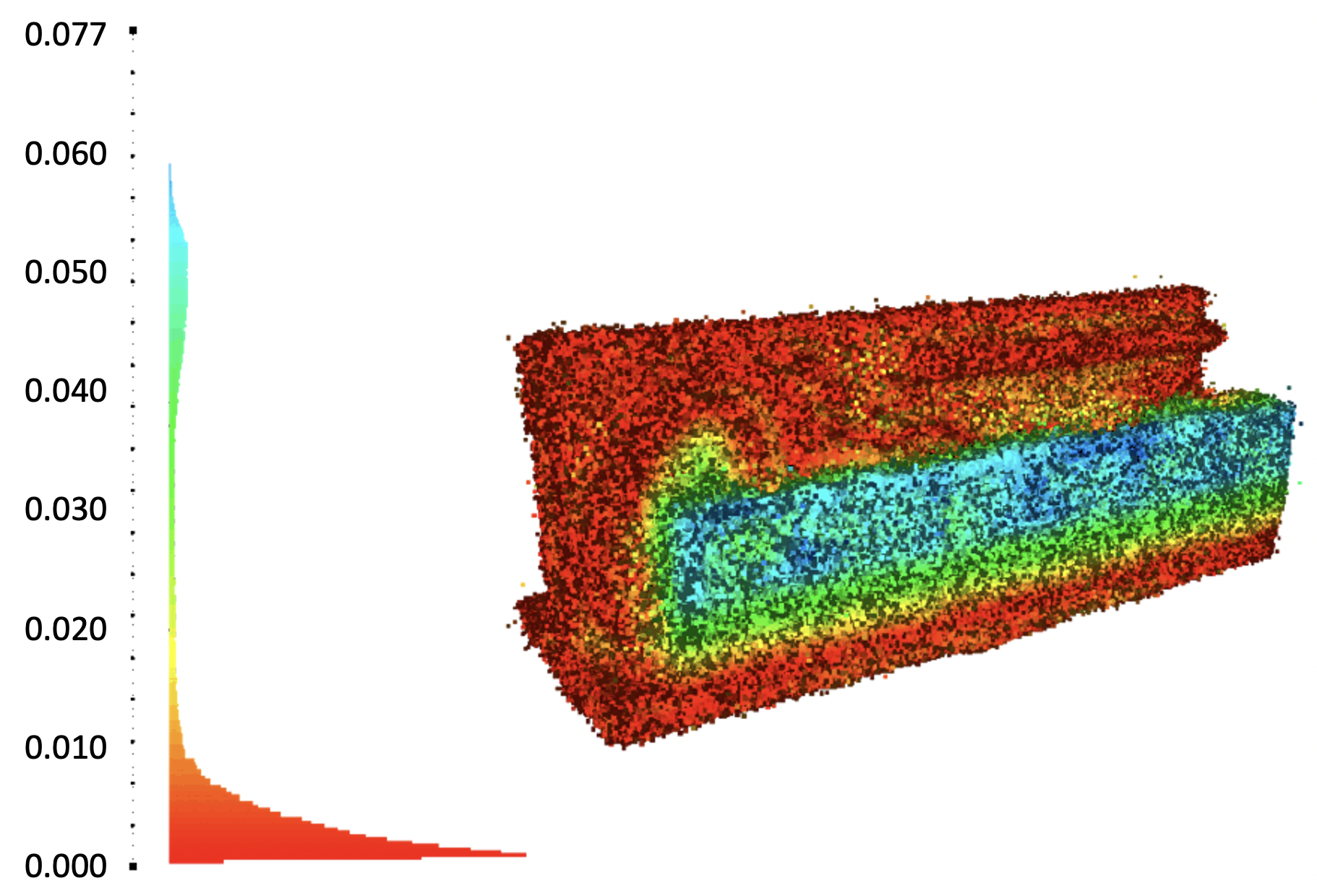}
    \caption{Anomaly detected}
\end{subfigure}
\hfill
\begin{subfigure}{0.238\textwidth}
    \centering
    \includegraphics[width=\linewidth]{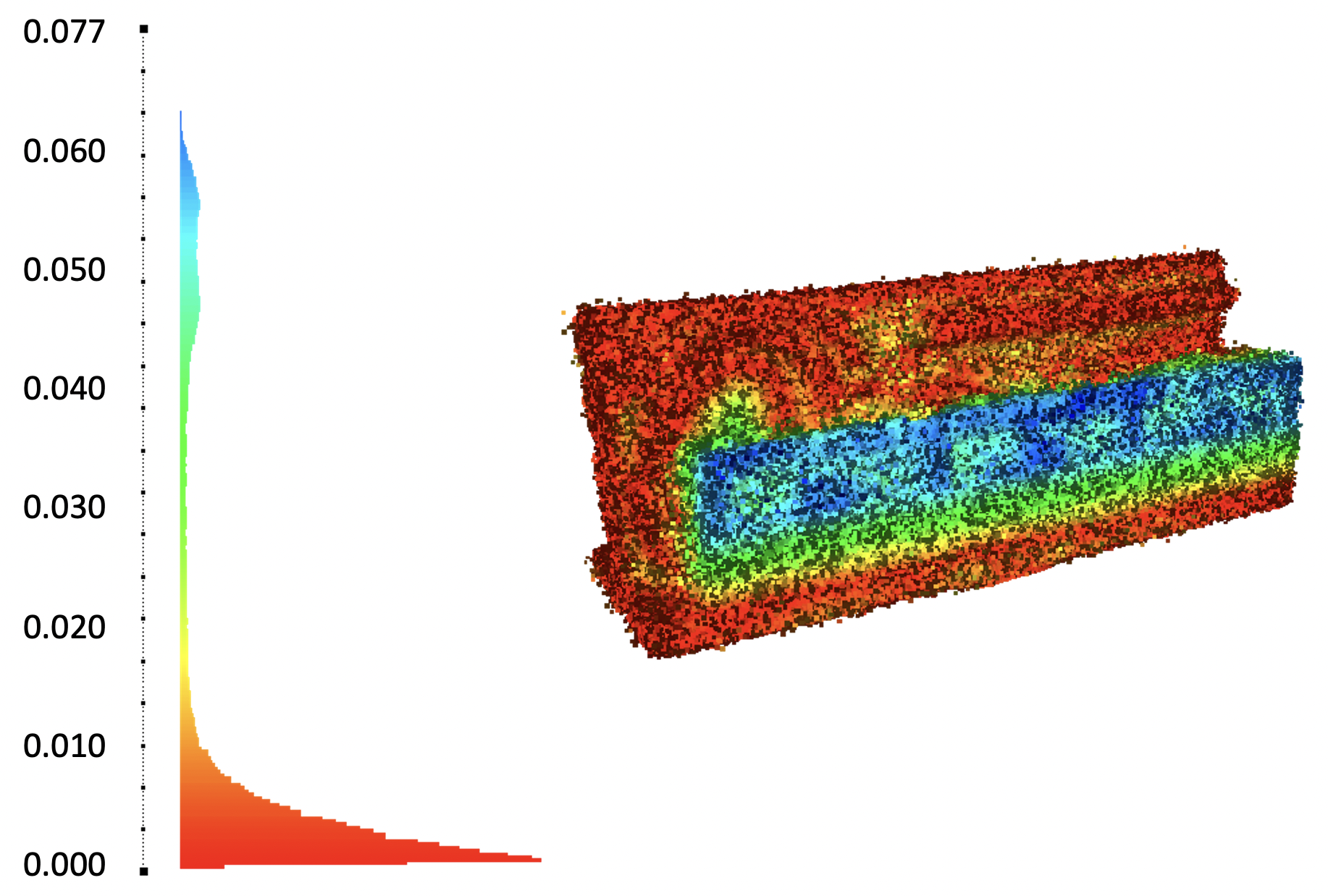}
    \caption{Location-aware detection}
\end{subfigure}
\vspace{-0.5mm} 
\caption{Demonstration of anomaly detection capabilities using single-UAV setup and Nerfacto}
\label{fig:anomaly_nerfacto}
\vspace{-5mm} 
\end{figure*}

As depicted in Figures~\ref{fig:reconstruction_1uav} and~\ref{fig:reconstruction_2uav}, the usage of \ac{GenAI} benefits the 3D reconstruction accuracy, density, and quality.
Focusing first on the metrics comparing the quality of \ac{3D} reconstruction to the input images, Splatfacto yields the best performance. 
Specifically, as visible in Table~\ref{tab:results_reconstruction}, Splatfacto outperforms Nerfacto and instant-ngp in terms of the \ac{SSIM} and latency metrics.
In addition, it visually yields the best quality of reconstruction when compared to the alternative approaches, as observable in Figures~\ref{fig:reconstruction_1uav} and~\ref{fig:reconstruction_2uav}.
Based on this observation, we argue that Splatfacto can provide the best visual quality of reconstruction, which, for the considered small scale objects, can be captured by utilizing the SSIM and latency metrics.

Table~\ref{tab:results_anomaly_detection} presents the performance derived for the reference and the reconstructed object with anomaly. The aim is to assess the consistently of the metrics for different objects for reconstruction, yet with comparable physical sizes.
As visible by comparing the single-UAV setup performance in Table~\ref{tab:results_reconstruction} with the corresponding performance of different approaches for a slightly modified object in Table~\ref{tab:results_anomaly_detection}, Nerfacto provides the most stable performance in relation to the object shapes. 
This is a desired feature for anomaly detection applications, in which different objects should be reconstructed with comparable accuracy, which is a precursor for anomaly detection.

An example spatial detection of an anomaly is depicted in Figure~\ref{fig:anomaly_nerfacto} for the Nerfacto model.
The anomaly consisted of a set of boxes covering the lower part of the letters on one side of the object, as depicted in Figure~\ref{fig:anomaly_nerfacto}a.
The difference $\Delta HD$ between the \ac{HD} distances of the object with and without the anomaly can be used for detecting the anomaly and its magnitude, as shown in the figure.

%!TEX root = ieee_pimrc_main.tex

%%%%%%%%%%%%%%%%%%%%%%%%%%%%%%%%%%%%%%%%%%%%%%%%%%%%%%%%%%%%%%%%%%%%%%%%%%%%%%%%%%%%%%%%%
%%%%%%%%%%%%%%%%%%%%%%%%%%%%%%%%%%%%%%%%%%%%%%%%%%%%%%%%%%%%%%%%%%%%%%%%%%%%%%%%%%%%%%%%%
%% Conclusions
%%%%%%%%%%%%%%%%%%%%%%%%%%%%%%%%%%%%%%%%%%%%%%%%%%%%%%%%%%%%%%%%%%%%%%%%%%%%%%%%%%%%%%%%%
%%%%%%%%%%%%%%%%%%%%%%%%%%%%%%%%%%%%%%%%%%%%%%%%%%%%%%%%%%%%%%%%%%%%%%%%%%%%%%%%%%%%%%%%%
\section{Conclusions and Future Efforts}
\label{sec:conclusion}

The integration of \ac{GenAI} with small \ac{UAV}-based systems for 3D digital reconstruction has proven to be an effective approach to addressing the limitations posed by such \acp{UAV}.
The utilization of \ac{GenAI} models significantly enhances the accuracy and realism of the reconstructions, even with the constraints in sensor quality. Future research can explore several areas to build upon this work to refine the technology.

One promising avenue is the incorporation of \ac{RGB} cameras, which would enhance the richness and detail of the captured images compared to the currently leveraged black-and-white ones. The inclusion of color data would likely improve the quality and realism of 3D reconstructions, particularly when dealing with complex textures or surfaces.
Improving the localization accuracy of small \acp{UAV} presents another opportunity. Current systems suffer from tracking inaccuracies, which can negatively affect the precision of 3D reconstructions. Future work should focus on refining localization techniques to mitigate these issues and ensure that data is captured with greater spatial accuracy.

Additionally, future assessments should evaluate the performance for larger objects to confirm its general applicability for various scales of reconstruction. A methodology that adjusts the small \ac{UAV} system based on the size of the object could be developed. This would involve optimizing parameters like \ac{UAV} positioning, flight patterns, and data collection rates in response to different object sizes, ensuring scalability while maintaining accuracy.
There is a potential to optimize 3D anomaly detection within these systems. The use of alternative distance metrics, such as mean instead of the currently utilized maximum distance, could lead to more precise and reliable detection. Further research could enhance the system’s ability to detect subtle discrepancies in reconstructed models, thereby expanding its utility to a broader range of applications.

\section*{Acknowledgments}
This work was supported by the European Union's Horizon Europe's programme (grants nº 101139161 - INSTINCT and nº 101192521 - MultiX projects). This work also received support from MCIN/AEI/10.13039/501100011033/FEDER/UE HoloMit 2.0 project (nº PID2021-126551OB-C21) and by the CERCA Programme of the Generalitat de Catalunya.

% \renewcommand{\bibfont}{\footnotesize}
% \printbibliography
\bibliographystyle{ieeetr}  % or another bibtex-compatible style
\bibliography{biblio}

\begin{thebibliography}{10}

\bibitem{nex2022uav}
F.~Nex, C.~Armenakis, M.~Cramer, D.~A. Cucci, M.~Gerke, E.~Honkavaara, A.~Kukko, C.~Persello, and J.~Skaloud, ``Uav in the advent of the twenties: Where we stand and what is next,'' {\em ISPRS journal of photogrammetry and remote sensing}, vol.~184, pp.~215--242, 2022.

\bibitem{floreano2015science}
D.~Floreano and R.~Wood, ``Science, technology and future of small autonomous drones,'' {\em Nature}, vol.~521, no.~7553, pp.~460--466, 2015.

\bibitem{muller2022instant}
T.~M{\"u}ller, A.~Evans, C.~Schied, and A.~Keller, ``Instant neural graphics primitives with a multiresolution hash encoding,'' {\em ACM transactions on graphics (TOG)}, vol.~41, no.~4, pp.~1--15, 2022.

\bibitem{nerfstudio}
M.~Tancik, E.~Weber, E.~Ng, R.~Li, B.~Yi, T.~Wang, A.~Kristoffersen, J.~Austin, K.~Salahi, A.~Ahuja, {\em et~al.}, ``Nerfstudio: A modular framework for neural radiance field development,'' in {\em ACM SIGGRAPH 2023 Conference Proceedings}, pp.~1--12, 2023.

\bibitem{kerbl3Dgaussians}
B.~Kerbl, G.~Kopanas, T.~Leimk{\"u}hler, and G.~Drettakis, ``3d gaussian splatting for real-time radiance field rendering,'' {\em ACM Transactions on Graphics}, vol.~42, no.~4, 2023.

\bibitem{correll2017wireless}
N.~Correll, P.~Dutta, R.~Han, and K.~Pister, ``Wireless robotic materials,'' in {\em Proceedings of the 15th ACM Conference on Embedded Network Sensor Systems}, pp.~1--6, 2017.

\bibitem{lemic2021survey}
F.~Lemic {\em et~al.}, ``Survey on terahertz nanocommunication and networking: A top-down perspective,'' {\em IEEE Journal on Selected Areas in Communications}, vol.~39, no.~6, pp.~1506--1543, 2021.

\bibitem{schenk2005introduction}
T.~Schenk, ``Introduction to photogrammetry,'' {\em The Ohio State University, Columbus}, vol.~106, no.~1, 2005.

\bibitem{kovanivc2023review}
L.~Kovani{\v{c}}, B.~Topitzer, P.~Petovsk{\`y}, P.~Bli{\v{s}}tan, M.~B. Gergelov{\'a}, and M.~Bli{\v{s}}tanov{\'a}, ``Review of photogrammetric and lidar applications of uav,'' {\em Applied Sciences}, vol.~13, no.~11, p.~6732, 2023.

\bibitem{kamencay2012improved}
P.~Kamencay, M.~Breznan, R.~Jarina, P.~Lukac, and M.~Zachariasova, ``Improved depth map estimation from stereo images based on hybrid method.,'' {\em Radioengineering}, vol.~21, no.~1, 2012.

\bibitem{gao2022nerf}
K.~Gao, Y.~Gao, H.~He, D.~Lu, L.~Xu, and J.~Li, ``Nerf: Neural radiance field in 3d vision, a comprehensive review,'' {\em arXiv:2210.00379}, 2022.

\bibitem{mildenhall2021nerf}
B.~Mildenhall, P.~P. Srinivasan, M.~Tancik, {\em et~al.}, ``Nerf: Representing scenes as neural radiance fields for view synthesis,'' {\em Communications of the ACM}, vol.~65, no.~1, pp.~99--106, 2021.

\bibitem{mendes2022small}
K.~Mendes {\em et~al.}, ``Small uavs-supported autonomous generation of fine-grained 3d indoor radio environmental maps,'' in {\em Distributed Computing Systems Workshops (ICDCSW)}, pp.~296--301, IEEE, 2022.

\bibitem{derpanis2010overview}
K.~G. Derpanis, ``Overview of the ransac algorithm,'' {\em Image Rochester NY}, vol.~4, no.~1, pp.~2--3, 2010.

\end{thebibliography}

\end{document}